\documentclass[prl,twocolumn,amsmath,amssymb]{revtex4}

\usepackage{graphicx}
\usepackage{dcolumn}
\usepackage{bm}
\usepackage{mathrsfs}
\usepackage{subfigure}
\usepackage{natbib}
\usepackage{amsmath}
\usepackage{amssymb}
\usepackage{Jonasmacros}
\usepackage{mathptmx}
\usepackage{txfonts}
\usepackage{ifsym}
\usepackage{pifont}

\usepackage[usenames,dvipsnames,svgnames]{xcolor}
\usepackage[colorlinks,citecolor=red,linkcolor=magenta]{hyperref}

\begin{document}
\date{\today}

\title{Vibrational Origin of Exchange Splitting and Chirality Induced Spin Selectivity}

\author{J. Fransson}
\email{Jonas.Fransson@physics.uu.se}
\affiliation{Department of Physics and Astronomy, Box 516, 75120, Uppsala University, Uppsala, Sweden}

\begin{abstract}
Electron exchange and correlations emerging from the coupling between ionic vibrations and electrons are addressed. Spin-dependent electron-phonon coupling originates from the spin-orbit interaction, and it is shown that such electron-phonon coupling introduces exchange splitting between the spin channels in the structure. By application of these results to a model for a chiral molecular structure mounted between metallic leads, the chirality induced spin selectivity is found to become several tens of percents using experimentally feasible parameters.
\end{abstract}

\maketitle

Chirality induced spin selectivity is an intriguing phenomenon that, to our knowledge, rests on a foundation of structural chirality, spin-orbit interactions, and strongly non-equilibrium conditions. The effect is a measure of the response to changes in the magnetic environment coupled, and the phenomenology refers back to the experimental observations of a substantial changes in the charge current amplitude through chiral molecules, upon changes in the external magnetic conditions \cite{Science.283.814,Science.331.894}.
Chirality induced spin selectivity has been shown to not be limited to multi-stranded helical structures, such as, double stranded DNA molecules \cite{NanoLett.11.4652} and bacteriorhodopsin \cite{PNAS.110.14872}, but has also been observed in, for example, various types of peptides \cite{NatComm.7.10744,NatComm.8.14567,AdvMat.30.1707390,JPhysChemLett.10.1139,JPhysChemLett} and polyalanines \cite{NatComm.4.2256,NanoLett.14.6042,Small.15.1804557}, and recently also in helicene \cite{AdvMater.28.1957,JPhysChemLett.9.2025}.

On the theoretical side, different approaches can be categorized into continuum models \cite{JChemPhys.131.014707,EPL.99.17006,JPCM.26.015008,PhysRevB.88.165409,JChemPhys.142.194308,PhysRevE.98.052221,PhysRevB.99.024418,NJP.20.043055,JPhysChemC.123.17043}, tight-binding descriptions \cite{PhysRevB.85.081404(R),PhysRevLett.108.218102,JPhysChemC.117.13730,PhysRevB.93.075407,PhysRevB.93.155436,ChemPhys.477.61,NanoLett.19.5253}, \emph{ab initio} simulations using, for example, density functional theory \cite{JPhysChemLett.9.5453,JPhysChemLett.9.5753,chemrxiv.8325248}, and multiple scattering off magnetic impurities \cite{arXiv:2020.04511}. Thus far, the success of the modeling has been limited to qualitative aspects while quantitative ones yet appear to be out of reach. Moreover, a major deficit with these approaches is their reliance on unrealistic spin-orbit interaction parameters. The source of this issue is likely to be found in the single-electron nature of these models.


Efforts to include electron correlations have been limited to the study in Ref. \cite{JPhysChemLett.10.7126}. Here, an on-site Coulomb interaction was added on each ionic site in a molecular chain, described by a generalized form of the Kane-Mele model \cite{PhysRevLett.61.2015,PhysRevLett.95.146802,PhysRevLett.95.226801}, including the spin-orbit interaction and adapted to a finite helical molecular structure. Using experimentally viable spin-orbit interaction parameters, it was shown that the exchange splitting between the spin channels that was introduced by the Coulomb interaction, supports a chirality induced spin selectivity of up to ten percents. The exchange splitting is, hence, a source for a substantial nonequivalence between the spin channels, which is maintained under reversal of the magnetic environment.

Another source for electron correlations pertaining to molecular structures is generated by molecular vibrations, or, phonons, that couple to, and distort, the electronic structure. The electron-phonon coupling may introduce an attractive electron-electron interaction, which can be shown by employing the canonical transformation proposed by Lang and Firsov \cite{ZhEkspTeorFiz.43.1843}. However, most theoretical studies concerned with electron-phonon coupling omit the analogous coupling to the spin degrees of freedom of the electrons, which can be justified whenever the spin-orbit interaction is negligible. The question if electron-phonon interactions make a difference in solids with non-negligible spin-orbit coupling has been addressed in the context of ultra-fast demagnetization \cite{NewJPhys.18.023012}, nevertheless, merely as a source for decoherence. Whether electron-phonon interactions is a pertinent mechanism for chirality induced spin selectivity is, however, an open question.

The results presented in this Letter shows that the cooperation of electron-phonon and spin-orbit interactions generates an exchange splitting between the spin-channels, which is viable for chirality induced spin selectivity. To this end, a mechanism for spin-dependent electron-phonon coupling is introduced and discussed. This mechanism is applied to a model for a general lattice structure, in which the emergence of a vibrationally induced spin splitting of the electron energies is stressed. This result is novel and should open for new ways to address high temperature magnetism. Finally, these findings are carried over to a model for a chiral molecule \cite{JPhysChemLett.10.7126}, which is used to show strong chirality induced spin selectivity, tens of percents, using experimentally feasible input parameters.



First, consider the mechanism for a coupling between the electron spin and the ionic vibrational degrees of freedom through the spin-orbit interaction. The single electron Hamiltonian is in the Schr\"odinger picture given by
\begin{align}
H=&
	\frac{p^2}{2m_e}
	+
	V(\bfr)
	+
	\frac{\xi}{2}
	\Bigl(
		\nabla V(\bfr)\times\bfp
	\Bigr)
	\cdot\bfsigma
	,
\label{eq-seHam}
\end{align}
where $m_e$ is the electron mass, $V(\bfr)$ is the effective confinement potential, and $\xi=\hbar/4m_e^2c^2$. Here, the operator $\bfp$ acts on everything to its right whereas $\nabla$ acts only on the component directly adjacent to its right, such that, $\nabla V\times\bfp=(\nabla V)\times\bfp$.

The ions are located at the positions $\bfr_m$, such that the potential $V(\bfr)=\sum_mV(\bfr-\bfr_m)$. Assuming that the ions are not fixed in space but move about there respective equilibrium positions, such that $\bfr_m=\bfr_m^{(0)}+\bfQ_m$, where $\bfQ_m=\bfr_m-\bfr_m^{(0)}$, the Taylor expansion $V(\bfr)=V_0+\sum_m\bfQ_m\cdot\nabla V(\bfr)_{|\bfr\rightarrow\bfr_m^{(0)}}+\cdots$ provides the coupling between the electrons and lattice vibrations. Here, $V_0\equiv \sum_mV(\bfr_m^{(0)})$ represents the equilibrium potential of the static ionic positions. Effecting this expansion in the model, Eq. (\ref{eq-seHam}), retaining at most linear orders in the displacement variable $\bfQ$, results in that $H=H_0+H_1+H_2$, where
\begin{subequations}
\label{eq-epmodel}
\begin{align}
H_0=&
	\frac{p^2}{2m_e}
	+
	V_0
	+
	\frac{\xi}{2}\sum_m\bigg[\nabla V(\bfr-\bfr_m^{(0)})\times\bfp\biggr]\cdot\bfsigma
	,
\\
H_1=&
	\sum_m
		\Biggl(
			\bfQ_m\cdot\nabla V(\bfr-\bfr_m^{(0)})
			+
			\frac{\xi}{2}
			\biggl[\nabla\Bigl(\bfQ_m\cdot\nabla V(\bfr-\bfr_m^{(0)})\Bigr)\times\bfp\biggr]\cdot\bfsigma
		\Biggr)
	.
\label{eq-Hep}
\end{align}
\end{subequations}

The vibrational quantum operator $a_{m\mu}$ ($a^\dagger_{m\mu}$), which is introduced according to the standard definitions, annihilates (creates) a vibration at the site $m$ in the mode $\mu$. Then, the ionic displacement is written $\bfQ_m=\sum_\mu l_{m\mu}\bfepsilon_{m\mu}(a_{m\mu}+a^\dagger_{m\mu})$, where $l_{m\mu}=\sqrt{\hbar/2\rho\nu\omega_{m\mu}}$ defines a length scale in terms of the density of vibrations $\rho$, system volume $\nu$, and vibrational frequency $\omega_{m\mu}$, whereas $\bfepsilon_{m\mu}$ denotes the polarization vector.

A second quantized form of $H$ is obtained by employing the expansion $\cc{}(\bfr,t)=\sum_m\cc{m}(t)\phi_m\ket{\sigma}/\nu$, where $\phi_m=\phi_m(\bfr)$ is an eigenstate of $H_0$. Hence, the zero Hamiltonian is given by $\Hamil_0=\sum_m\psi^\dagger_m\bfE_m\psi_m$,
where the spinor $\psi_m=(\cs{m\up}\ \cs{m\down})^t$, whereas the on-site energy spectrum is defined by
\begin{align}
\bfE_m=&
	\int
		\phi_m^*
		H_0
		\phi_m
	\frac{d\bfr}{\nu}
\label{eq-spectrum}
	.
\end{align}
In the same fashion, the electron-phonon contributions to the model, $H_1$, can be written as
\begin{align}
\Hamil_\text{e-ph}=&
	\sum_{\stackrel{\scriptstyle mm'}{n\mu}}
		\psi^\dagger_m
		\Bigl(
			U_{mm'n\mu}
			+
			\bfJ_{mm'n\mu}\cdot\bfsigma
		\Bigr)
		\psi_{m'}
		\Bigl(
			a_{n\mu}+a^\dagger_{n\mu}
		\Bigr)
	,
\label{eq-Hep}
\end{align}
where
\begin{subequations}
\label{eq-UJ}
\begin{align}
U_{mm'n\mu}=&
	l_{n\mu}
	\sum_k
	\int
		\phi^*_m
			\bfepsilon_{\bfq\mu}\cdot\nabla V(\bfr-\bfr_k^{(0)})
		\phi_{m'}
	\frac{d\bfr}{\nu}
\label{eq-U}
	,
\\
\bfJ_{mm'n\mu}=&
	\frac{\xi l_{n\mu}}{2}
	\sum_k
	\int
		\phi^*_m
			\nabla
			\Bigl(
					\Bigl[\bfepsilon_{n\mu}\cdot\nabla V(\bfr-\bfr_k^{(0)})\Bigr]
					\times\bfp
			\Bigr)
		\phi_{m'}
	\frac{d\bfr}{\nu}
\label{eq-J}
	.
\end{align}
\end{subequations}

The derivation of $\Hamil_\text{e-ph}$ demonstrates the existence of a direct coupling between the electronic spin degrees of freedom and the ionic vibrations. In general, spin-orbit interaction requires a redefinition of what good spin quantum numbers means, c.f. $\Hamil_0$. However, when accompanied with the electron-phonon coupling, the spin-orbit interaction also provides an origin for spin-dependent electron-phonon interactions $\bfJ_{mm'n\mu}$. Hence, the spin-orbit and electron-phonon interactions combine into a viable spin-phonon coupling, Eq. ({\ref{eq-J}), in addition to the generic coupling $U_{mm'n\mu}$ between charge and vibrations, c.f., Eq. (\ref{eq-U}). This background serves as a justification for the succeeding discussion.

In order to stress a few implications emerging from the spin-dependent electron-phonon interactions, consider a dimer model. Each site ($m=1,2$) in the dimer vibrates with the frequency $\omega_m$ and comprises an electron level a the energy $\dote{m}$. The electronic and vibrational degrees of freedom are coupled through the potential $\bfU=U\sigma^0+J\sigma^z$, while the elastic spin-orbit coupling (third term in Eq. (\ref{eq-epmodel})), as well as the site dependences of the electron-phonon couplings are both negligible. The Hamiltonian model can, hence, be written as
\begin{align}
\Hamil=&
	\sum_m
		\Bigl(
			\dote{m}\psi^\dagger_m\psi_m
			+
			\omega_m a^\dagger_m a_m
		\Bigr)
	+
	\sum_{mn}
		\psi^\dagger_m
		\bfU
		\psi_m
		(a_n+a^\dagger_n)
	.
\label{eq-epspinmodel}
\end{align}
This model allows a for a transparent derivation of an electron-phonon generated spin-polarization, considered next.

Hence, the canonical transformation, $\widetilde\Hamil=e^S\Hamil e^{-S}$, with the generating operator \cite{ZhEkspTeorFiz.43.1843}
\begin{align}
S=&
	-\sum_{mn}
		\frac{1}{\omega_n}
		\psi^\dagger_m
		\bfU
		\psi_m
		(a_n-a^\dagger_n)
	,
\end{align}	
transform the electron and phonon operators according to
\begin{align}
\tilde{c}_{m\sigma}=
	\cc{m}e^{U_\sigma\sum_n(a_n-a_n^\dagger)/\omega_n}
	&,&
\tilde{a}_m=
	a_m
	-
	\frac{1}{\omega_m}
	\sum_n
		\psi^\dagger_n
		\bfU
		\psi_n
	,
\end{align}
respectively, where $U_\sigma=U+J\sigma^z_{\sigma\sigma}$, such that $\tilde{n}_{m\sigma}=\tilde{c}^\dagger_{m\sigma}\tilde{c}_{m\sigma}=\cdagger{m}\cc{m}=n_{m\sigma}$. Hence, the transformed model acquires the decoupled form
\begin{align}
\widetilde\Hamil=&
	\sum_m
		\Bigl(
			\dote{m}\psi^\dagger_m\psi_m
			+
			\omega_ma^\dagger_ma_m
		\Bigr)
	-
	\sum_n\frac{1}{\omega_n}
	\bigg(
		\sum_m
			\psi^\dagger_m\bfU\psi_m
	\biggr)^2
	.
\end{align}
In the following, the interaction parameters are redefined through $U\sqrt{\sum_m1/\omega_m}\rightarrow U$ and $J\sqrt{\sum_m1/\omega_m}\rightarrow J$.

Next, the interaction term $(\sum_m\psi^\dagger_m\bfU\psi_m)^2$ can be partitioned into three contributions $\hat{U}_i$, $i=1,2,3$, where
\begin{subequations}
\label{eq-U2}
\begin{align}
\hat{U}_1=&
	\sum_m
		\Bigl(
			(U^2+J^2)n_m
			+
			4UJs_m^z
		\Bigr)
	,
\label{eq-single}
\\
\hat{U}_2=&
	2(U^2-J^2)
	\sum_m
		n_{m\up}n_{m\down}
	,
\label{eq-onsite}
\\
\hat{U}_3=&
	2U^2n_1n_2
	+
	4UJ(n_1s_2^z+s_1^zn_2)
	+
	8J^2s_1^zs_2^z
	,
\label{eq-intersite}
\end{align}
\end{subequations}
where $n_m=\psi^\dagger_m\psi_m$ and $s_m^z=\psi^\dagger_m\sigma^z\psi_m/2$.
Hence, writing $\tilde{\dote{}}_m=\dote{m}-U^2-J^2$, the transformed model can be written as
\begin{align}
\widetilde\Hamil=&
	\sum_m
	\Bigl(
		\tilde{\dote{}}_mn_m
		-
		4JUs_m^z
		-
		2(U^2-J^2)n_{m\up}n_{m\down}
		+
		\omega_ma^\dagger_ma_m
	\Bigr)
\nonumber\\&
	-
	2U^2n_1n_2
	-
	4UJ(n_1s_2^z+n_2s_1^z)
	-
	8J^2s_1^zs_2^z
	.
\label{eq-tildeH}
\end{align}

The important observations to be made in Eqs. (\ref{eq-U2}) and (\ref{eq-tildeH}) are that the electron-phonon coupling has been converted into (i) a renormalized and spin-dependent single-electron energy, Eq. (\ref{eq-single}), and (ii) on-site and inter-site charging and exchange interactions, Eqs. (\ref{eq-onsite}) and (\ref{eq-intersite}), respectively. Hence, the one-electron states $\ket{m\sigma}=\cdagger{m}\ket{}$ are associated with the spin-dependent energies $E_{m\sigma}=\tilde{\dote{}}_m-2UJ\sigma^z_{\sigma\sigma}$, whereas the two-electron states $\ket{\sigma\sigma'}=\csdagger{2\sigma'}\cdagger{1}\ket{}$ are associated with the spin-dependent energies $E_{\sigma\sigma}=\sum_m\tilde{\varepsilon}_m-2U_\sigma^2-4UJ\sigma^z_{\sigma\sigma}$ and $E_{\up\down}=E_{\down\up}=\sum_m\tilde{\varepsilon}_m-2U_\up U_\down$. The remaining two-electron states $\ket{2m}\equiv\csdagger{m\down}\csdagger{m\up}\ket{}$ have the energies $E_{2m}=2\tilde{\dote{}}_m-2U_\up U_\down$. The three-electron states acquire a similar spin-split as the one-electron states, whereas the empty state and the four-electron state both are spin-independent.

In a, slightly, more realistic model, on- and inter-site electron-electron interactions should be included with associated parameters which, typically, are significantly larger than the induced parameters $U$ and $J$. In such a consideration, the effective correlation energies are, therefore, merely renormalized by the electron-phonon interaction. The one-electron (as well as the two- and three-electron) energies would, nonetheless, remain spin-polarized as is indicated by the second contribution in Eq. (\ref{eq-single}). It is also worth pointing out that the results are not limited to dimers but can be straightforwardly generalized to assemblies with any number of ions.

The results from the above model, Eqs. (\ref{eq-epspinmodel}) -- (\ref{eq-tildeH}), emphasize that, when accompanied with spin-orbit coupling, the electron-phonon interactions should be interpreted as a source for exchange splitting between the spin channels.
This result becomes especially important in the context of chirality induced spin selectivity, where a finite exchange splitting is crucial for the emergence of a large effect when using realistic parameters \cite{JPhysChemLett.10.7126}. The existence of the exchange does, however, not have to originate from electronic Coulomb interactions.

Hence, the final part of this Letter is devoted to the electron-phonon induced spin selectivity in a chiral chain of ions comprising $\mathbb{M}$ sites, modeled as
\begin{align}
\Hamil_\text{mol}=&
	\sum_{m=1}^{\mathbb{M}}
		\dote{m}\psi^\dagger_m\psi_m
	+
	\sum_{\nu=1}^\mathbb{M}\omega_\nu a^\dagger_\nu a_\nu
\nonumber\\&
	-
	\sum_{m=1}^{\mathbb{M}-1}
		\Bigl(
			\psi^\dagger_m\psi_{m+1}
			+
			H.c.
		\Bigr)
		\biggl(
			t_0
			+
			t_1\sum_\nu(a_\nu+a^\dagger_\nu)
		\biggr)
\label{eq-chiralH}\\&
	+
	\sum_{m=1}^{\mathbb{M}-2}
		\Bigl(
			i\psi^\dagger_m\bfv_m^{(+)}\cdot\bfsigma\psi_{m+2}
			+
			H.c.
		\Bigr)
		\biggl(
			\lambda_0
			+
			\lambda_1\sum_\nu(a_\nu+a^\dagger_\nu)
		\biggr)
	.
\nonumber
\end{align}
Here, the molecule is described by a set of single electron level described by $\dote{m}\psi^\dagger_m\psi_m$, first term, where $\dote{m}$ denotes the level energy, distributed along the spatial coordinates $\bfr_m=(a\cos\varphi_m,a\sin\varphi_m,(m-1)c/(\mathbb{M}-1))$, $\varphi=(m-1)2\pi/N$, $m=1,\ldots,\mathbb{M}$. Here, $a$ and $c$ define the radius and length, respectively, of the helical structure, whereas $\mathbb{M}=MN$ is the total number of sites in which product $M$ and $N$ denote the number of laps and ions per lap, respectively. Nearest-neighboring sites, second line in $\Hamil_\text{mol}$, interact via direct hopping, rate $t_0$, and electron-phonon assisted hopping, rate $t_1$. Similarly, the spin-orbit coupling is picked up between next-nearest neighbor sites, last line in $\Hamil_\text{mol}$, through processes of the type $i\psi_m^\dagger\bfv_m^{(s)}\cdot\bfsigma\psi_{m+2s}$, $s=\pm1$, where $\lambda_0$ and  $\lambda_1$ denote the direct and electron-phonon assisted spin-orbit interaction parameters, respectively. The vector $\bfv_m^{(s)}=\hat\bfd_{m+s}\times\hat\bfd_{m+2s}$ defines the chirality of the helical molecule in terms of the unit vectors $\hat{\bfd}_{m+s}=(\bfr_m-\bfr_{m+s})/|\bfr_m-\bfr_{m+s}|$. The electrons are coupled to the $\mathbb{M}$ vibrational normal modes $\omega_\nu$, which are represented by the phonon operators $a_\nu$ and $a^\dagger_\nu$.

The transport properties are captured by mounting the molecule in the junction between a ferromagnetic and a simple metallic lead. Tunneling between the lead $\chi=L,R$, and the adjacent ionic site introduces a spin-resolved level broadening through the coupling parameter $\bfGamma^\chi_\sigma=\Gamma^\chi(\sigma^0+p_\chi\sigma^z)$. Here, $|p_\chi|\leq1$ denotes the spin-polarization, whereas $\Gamma^\chi=2\pi\sum_{\bfk\in\chi}|t_\chi|^2\rho_\chi(\dote{\bfk})$ accounts for the tunneling rate $t_\chi$ and the density of electron states $\rho_\chi(\dote{\bfk})$ in the lead $\chi$.

We relate the properties of the electronic structure and transport to the single electron Green function $\bfG_{mn}(z)=\av{\inner{\psi_m}{\psi^\dagger_n}}(z)$, $\bfG_m\equiv\bfG_{mm}$, through the density of electron states $\rho_m(\omega)=i~{\rm sp}[\bfG^>_{mm}(\omega)-\bfG^<_{mm}(\omega)]/2\pi$ and charge current
\begin{align}
J_\sigma=&
	\frac{ie}{h}
	{\rm sp}
	\int
		\bfGamma^L_\sigma
		\Bigl(
			f_L(\omega)\bfG^>_1(\bfGamma^L_\sigma;\omega)
			+
			f_L(-\omega)\bfG^<_1(\bfGamma^L_\sigma;\omega)
		\Bigr)
	d\omega
	,
\end{align}
respectively, where ${\rm sp}$ denotes the trace over spin 1/2 space, whereas $f_\chi(\omega)=f(\omega-\mu_\chi)$ is the Fermi-Dirac distribution function at the chemical potential $\mu_\chi$. The notation $\bfG^{</>}_1(\bfGamma^L_\sigma;\omega)$ indicates that the calculated properties depend on the spin-polarization $p_L$ in the left lead through the coupling parameter $\bfGamma^L_\sigma$. All calculations are made using non-equilibrium Green functions $\bfG^{</>}_{mn}(\omega)$.

The equation of motion for the Green function $\bfG_{mn}=\bfG_{mn}(z)$ can be written on the form
\begin{align}
\Bigl(
	z&
	-
	E_m
	\Bigr)
	\bfG_{mn}
	-
	\sum_{s=\pm1}
		\Biggl\{
			-t_0\bfG_{m+sn}
			+i\lambda_0\bfv_m^{(s)}\cdot\bfsigma\bfG_{m+2sn}
\nonumber\\&
			+
			\sum_{s'=\pm1}
				\bfSigma_m
				\biggl(
					t_1^2\bfG_{m+s+s'n}
					-\lambda_1^2\bfv_m^{(s)}\cdot\bfsigma\bfv_{m+2s}^{(s')}\cdot\bfsigma\bfG_{m+2(s+s')n}
\nonumber\\&
					-it_1\lambda_1\bfsigma\cdot
					\Bigl(
						\bfv_m^{(s)}\bfG_{m+2s+s'n}
						+
						\bfv_{m+s}^{(s')}\bfG_{m+s+2s'n}
					\Bigr)
				\biggr)
		\Biggr\}
	=
	\delta_{mn}
	.
\end{align}
Here, $E_1=\dote{1}-i\bfGamma^L_\sigma/2$, $E_m=\dote{m}$, $2\leq m\leq\mathbb{M}-1$, $E_\mathbb{M}=\dote{\mathbb{M}}-i\bfGamma^R/2$, and $\bfG_{mn}=0$ for $m,n\notin\{1,2,\ldots,\mathbb{M}\}$. The self-energy $\bfSigma_m=\bfSigma_m(z)$ includes the simplest non-trivial electron-phonon interactions, given by
\begin{align}
\bfSigma_m(z)=&
	\sum_\nu
		\biggl(
			\frac{n_B(\omega_\nu)+1-f(\dote{m})}{z-\omega_\nu-\dote{m}}
			+
			\frac{n_B(\omega_\nu)+f(\dote{m})}{z+\omega_\nu-\dote{m}}
		\biggr)
	,
\label{eq-Sigma}
\end{align}
where $n_B(\omega)$ is the Bose-Einstein distribution function.
As the exchange splitting generated by the ionic vibrations is an intrinsic property of the structure, the employed approach is justified since it captures the main effect of the electron-phonon coupling. Hence, despite non-equilibrium conditions may modify the exchange splitting, the gross effect of the electron-phonon interactions is captured by using the equilibrium self-energy.

\begin{figure}[t]
\begin{center}
\includegraphics[width=\columnwidth]{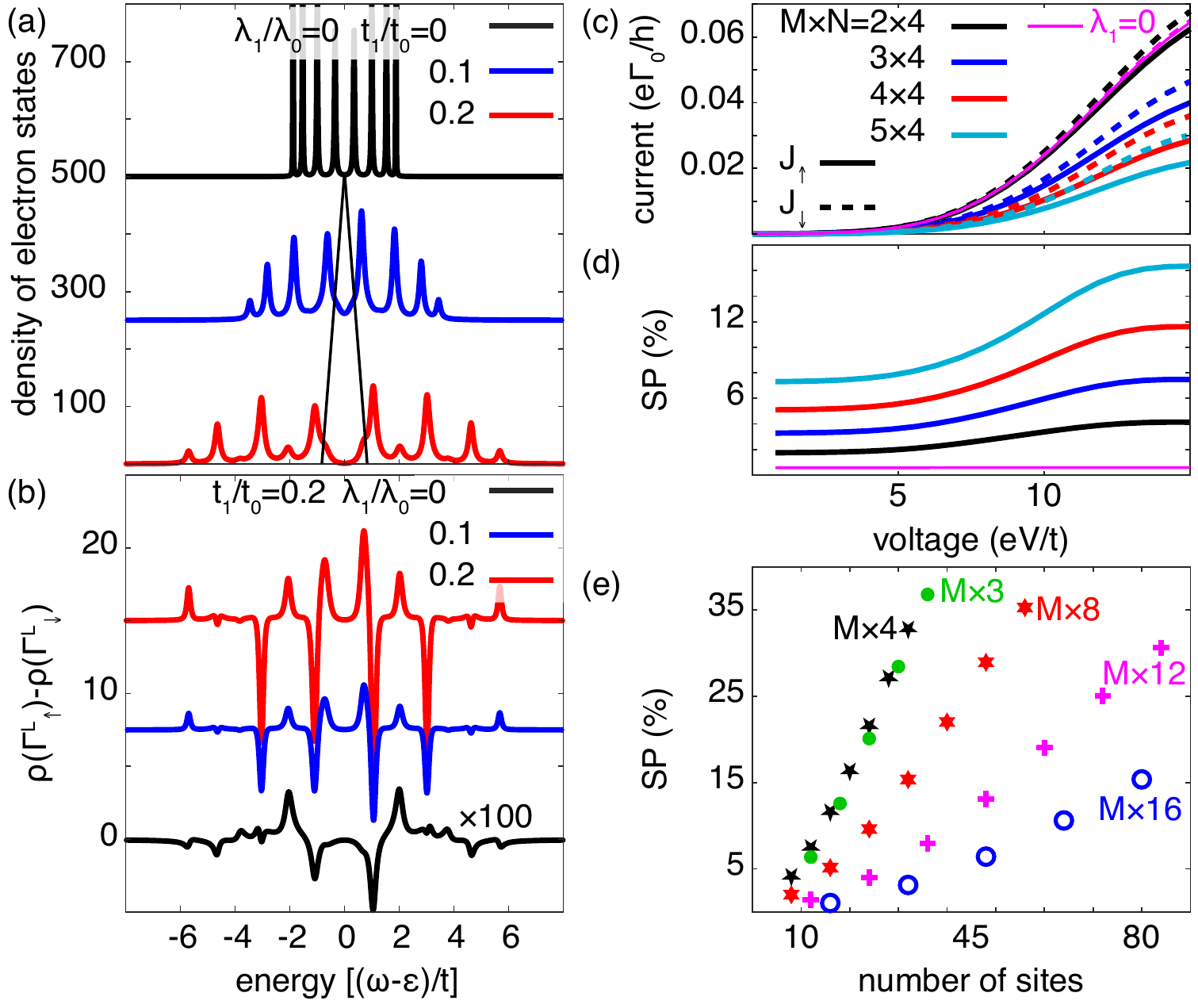}
\end{center}
\caption{(Color online)
(a) Density of electron states of a chiral molecule ($M\times N=2\times4$) for increasing $t_1=0$ (black), $0.1t_0$ (blue), and $0.2t_0$ (red), using $\lambda_1=0$.
(b) Difference between the density of electron states for spin-polarization $\up$ ($p_L=0.5$) and  $\down$ ($p_L=-0.5$), for $t_1=0.2t_0$, and $\lambda_1=0$ (black), $0.1\lambda_0$ (blue), and $0.2\lambda_0$ (red).
(c) Charge currents and (d) corresponding spin selectivities, for spin-polarization $\up$ (solid) and $\down$ (dashed), for structures with $2\times4$ (black), $3\times4$ (blue), $4\times4$ (red), and $5\times4$ (cyan) sites. Currents in absence of vibrations (magenta) are shown for reference.
(e) Spin selectivity at $eV/t_0=15$ for increasing number of laps ($M$) and number of sites per lap ($N$), using $t_1=0.1t_0$, $\lambda_1=0.1\lambda_0$.
Other parameters are $t_0=40$ meV, $\lambda_0=t_0/40$, $\dote{}-\dote{F}=-6t_0$, $\Gamma_0=1/\tau_\text{ph}=t_0/4$, $p_L=\pm0.5$, $p_R=0$, $\omega_\nu/t=\{0.010,\ 1.4,\ 2.9,\ 4.3,\ 5.7,\ 7.1,\ 8.6,\ 10.0\}$, and $T=300$ K.
Results in (a) and (b) were made for a $2\times4$ structure.
An intrinsic broadening $1/\tau_\text{ph}=t_0/4$ has been used in the phonon propagation in order to smoothen the electronic densities.
}
\label{fig-Results}
\end{figure}

The results form Eq. (\ref{eq-tildeH}) indicates that a correlation gap opens under finite electron-phonon interaction. This is corroborated in the model for the chiral structure, see Fig. \ref{fig-Results} (a), where the density of electron states for a chiral molecule ($M\times N=2\times4$) is plotted for increasing spin-independent electron-phonon coupling $t_0$, keeping $\lambda_0=0$. See the figure caption for other parameters. The opening of the correlation gap and accompanied splitting of the density into two bands is indicated with the faint lines around zero.

Opening of a correlation gap is, however, not crucial for spin selectivity. The reason is that spin selectivity leads to a changed charge current through the structure upon switching the direction of the magnetic environment; here, the spin-polarization in the left lead. However, the vibrationally generated correlation gap is not associated with a broken time-reversal symmetry, which can be seen in Fig. \ref{fig-Results} (b), where the bottom trace shows that the difference of the density of electron states for the two spin-polarizations $p_L=\pm0.5$ is essentially negligible. The existing difference is due to higher order coupling between vibrationally assisted tunneling ($t_1$) and spin-orbit induced tunneling ($\lambda_0$). The implication is, hence, that the corresponding currents are roughly equal and the generated spin selectivity is minute, see Fig. \ref{fig-Results} (c), (d) (faint/magenta).

Inclusion of the spin-dependent component of the electron-phonon interaction leads, as suggested by Eq. (\ref{eq-tildeH}), to a finite exchange splitting between the spin channels. Indeed, already for weak vibrationally assisted spin-orbit couplings, the asymmetry of the electronic density with respect to the spin-polarization $p_L$ becomes enhanced by orders of magnitude, see middle and top traces in Fig. \ref{fig-Results} (b). It is, therefore, anticipated that the spin selectivity should be sizable in presence of coupling between the spin-orbit and electron-phonon interactions. The plots in Fig. \ref{fig-Results} (c), (d), display the polarization ($p_L$) dependence of the charge current for $\lambda_1/\lambda_0=0.1$, and the corresponding spin selectivity $\text{SP}=100|(J_\up-J_\down)/(J_\up+J_\down)|$, for chiral structures with $M\times4$, $M=2$, 3, 4, 5, number of sites. The results clearly demonstrate the emergence of a significant spin selectivity in presence of a finite spin-dependent electron-phonon coupling $\lambda_1$.
It is, moreover, important to notice that the overall current reduces with increasing number of sites while the spin selectivity simultaneously increases. The latter property is expected since the number of laps in the helix tends to accumulate and, hence, increase the effect of the induced exchange splitting.

The plots in Fig. \ref{fig-Results} (e), finally, shows the dependence of the spin selectivity as function of the product $M\times N$, indicating a stronger spin selectivity the fewer the sites per lap in the helix. This is expected since the chirality vector $\bfv_m^{(s)}\sim\sin\varphi_m$, where $\varphi_m$ is the angle between the vectors comprised in $\bfv_m^{(s)}$. In the limit of infinitely many sites per lap, the chirality vanishes since the sites comprised in the product lie on the same straight line, which leads that $\varphi_m\rightarrow0$. For decreasing number of sites per lap, $\varphi_m$ becomes finite and approaches $\pi$ in the limit of two sites per lap.

In the presented calculations, the vibrational modes are equidistantly distributed in the range between $\{\omega_\nu/t_0\}=\{0.01,\ldots,10\}$, where the reciprocal of the lowest mode, in combination with $t_1$, sets the energy scale for the vibrationally induced correlation gap, c.f., $\sum_mU/\omega_m$ in Eq. (\ref{eq-tildeH}). Low frequency modes are abundant in organic molecules and, therefore, the used spectrum should be considered as realistic. Moreover, the temperature has been set to 300 K in order to comply with experimental conditions, while the parameters $t_0=40$ meV and $\lambda_0=t_0/40$ correspond to realistic numbers.

A weakness of the present study is the absence of Coulomb interactions as well as the equilibrium treatment of the self-energy in Eq. (\ref{eq-Sigma}). While the former issue may be considered, see \cite{JPhysChemLett.10.7126}, a full treatment, including both Coulomb and vibrational interactions, becomes numerically unattractive for structures with many sites. Concerning the latter issue, while it is doable to treat the self-energy in appropriate non-equilibrium framework, it does not change the qualitative outcome of the study but instead tends to complicate the calculations unnecessarily. By contrast, the main purpose of the present Letter is to point out the importance of the exchange splitting in the context of spin selectivity and that this splitting may emerge from vibrational sources.

In this Letter it has been shown that spin-dependent electron-phonon coupling has a realistic and sound foundation at the same level as the more commonly studied coupling between the electronic charge and lattice vibrations. The presence of both spin-independent and spin-dependent electron-phonon couplings leads to vibrationally induced exchange splitting between the spin-channels in the structure, a result which is novel and may open routes to addressing high temperature magnetism. These results were shown to have a great importance in the theoretical modeling and comprehension of chirality induced spin selectivity. Using experimentally realistic numbers, particularly for the spin-orbit interaction, the calculated spin selectivity reaches several tens of percents and is, therefore, comparable to experimental observations.

\acknowledgements
Support from Vetenskapsr\aa det, Stiftelsen Olle Engkvist Byggm\"astare, and Carl Tryggers Stiftelse is acknowledged.

\end{document}